\documentclass[aps,prl,twocolumn,preprintnumbers,10pt,showpacs,nohyper]{revtex4}
\usepackage{amsmath,bm}
\usepackage{multirow}

\usepackage{graphicx}
\usepackage{xspace}
\usepackage{epsfig}
\usepackage{color}

\newcommand{\NNLOJET}{NNLO\protect\scalebox{0.8}{JET}\xspace}

\def\nf{N_{F}}

\allowdisplaybreaks

\begin{document}

\preprint{IPPP/16/110,
MPP-2016-322}

\title{NNLO QCD predictions for single jet inclusive production at the LHC}

\author{J.\ Currie$^a$, E.W.N.\ Glover$^a$, J.\ Pires$^b$}
\affiliation{$^a$ Institute for Particle Physics 
Phenomenology, University of Durham, Durham DH1 3LE, England\\
$^b$ Max-Planck-Institut f\"ur Physik,
F\"ohringer Ring 6
D-80805 Munich, Germany}


\begin{abstract}

We report the first calculation of fully differential jet production at leading colour in all partonic channels at next-to-next-to 
leading order (NNLO) in perturbative QCD and compare to the available ATLAS 7 TeV
data. We discuss the size and shape of the perturbative corrections along with their associated scale variation
across a wide range in jet transverse momentum, $p_{T}$, and rapidity, $y$. We find significant effects, especially
at low $p_{T}$, and discuss the possible implications for Parton Distribution Function fits.

\end{abstract}

\maketitle

The Large Hadron Collider (LHC) is currently colliding protons at
centre of mass energies up to $\sqrt{s}=13$ TeV. The main goal
is to search the high energy frontier for signs of physics beyond the Standard Model.
 However, any searches for new physics are irreducibly dependent on how well
we understand the Standard Model and the collider environment of the LHC itself.

At the LHC the inclusive cross section for a given final-state can be calculated using the factorization formula,
\begin{eqnarray}
{\rm{d}}\sigma&=&\sum_{i,j}\int\frac{{\rm{d}}\xi_{1}}{\xi_{1}}\frac{{\rm{d}}\xi_{2}}{\xi_{2}}
f_{i}(\xi_{1},\mu_{F})f_{j}(\xi_{2},\mu_{F}){\rm{d}}\hat{\sigma}_{ij}
\end{eqnarray}
which is accurate up to non-pertubative hadronization corrections, typically of the order
 $\Lambda_{QCD}/Q$, where $Q$ is the hard scale in the scattering process. The partonic 
 cross section, ${\rm{d}}\hat{\sigma}_{ij}$, can be calculated as a perturbative series in the strong coupling, $\alpha_{s}$,
 and systematically improved by progressively including higher order terms in the series.
 It is also necessary to have a good understanding of the non-perturbative Parton Distribution Functions
 (PDF), $f_{i}(\xi,\mu_{F})$. The PDFs quantify the relative parton content of the proton carrying a 
 fraction, $\xi$, of the proton's momentum for a given factorization scale, $\mu_{F}$. 
 To calculate the cross section using this formula we need accurate determinations
 of the PDFs, $\alpha_{S}$ and the higher order terms in the perturbative expansion of the partonic
 cross section. 
  
Data from lepton-nucleon Deep Inelastic Scattering (DIS) experiments such as 
HERA~\cite{disbook} provide detailed information about the quark PDFs and have been used
to significantly constrain the uncertainties on these quantities. The inclusive cross section
in DIS involves the exchange of a virtual photon coupling to quarks at lowest order via the
 electroweak coupling constant. The electrical neutrality of the gluon means that the gluon PDF 
can only be constrained using specific final-states, such as heavy quarks or jets~\cite{disjet}, or indirectly
through DGLAP evolution of the flavour singlet distribution. 
In contrast, jet production at the Tevatron~\cite{cdf,d0} and LHC directly
probes the gluon PDF and is ${\cal{O}}(\alpha_{s}^{2})$ at leading order (LO).
 The single jet inclusive cross section has
been measured accurately by ATLAS~\cite{atlasjet,atlasjet2} and CMS~\cite{cmsjet} across
the large dynamical range of the LHC.

To take advantage of the available data we must be able to calculate observables
 with sufficient precision yet the cross section for producing jets
 is currently only known exactly at next-to leading order (NLO)~\cite{eks,jetrad,powheg2j,nlojet1,meks}
 and partially at NNLO~\cite{partial}. 
 The theoretical uncertainty in this observable, estimated from the dependence
 on unphysical scales, is the main limiting factor when
 determining parameters like $\alpha_{s}$ from jet data or consistently including this data
 in global fits for PDFs~\cite{ABM12,NNPDF,MMHT,CT14}. To improve on the status-quo it is clear that an accurate and
 precise determination of jet production at the LHC is needed and so in this letter
 we present the first calculation of the NNLO correction to jet production in perturbative
 QCD. Higher order corrections have the potential to change the size and shape of
 the cross section and also to reduce the residual scale dependence in a calculation;
 we discuss the extent to which this is true for the NNLO correction to the fully differential 
 single jet inclusive cross section.
  
 Predictions for jet production at NNLO accuracy require the relevant 
 tree-level~\cite{real}, one-loop~\cite{fiveg1l,2q3g1l,4q1g1l} and two-loop~\cite{twol,quarkgluon2l,fourq2l} 
 parton-level scattering amplitudes as well as a
 procedure for dealing with the infrared (IR) singularities present in both the phase space
 integrals and matrix elements, but which cancel in any IR safe physical observable.
 Several techniques have been developed for obtaining finite cross sections at NNLO
 for hadronic initial-states: antenna subtraction~\cite{antenna1,antenna2}, $q_{T}$-subtraction~\cite{qtsub}, 
$N$-jettiness subtraction~\cite{njettiness}, sector-improved residue subtraction~\cite{stripper},
sector decomposition~\cite{secdec} and projection to Born~\cite{projecttoborn}. We use
the antenna subtraction method, implemented in the parton-level event
generator, \NNLOJET~\cite{zjet,zjetproc}, to calculate the single jet inclusive cross section, fully differential
in the jet transverse momentum, $p_{T}$ and rapidity, $y$.

We include the leading colour contribution from all partonic subprocesses
in all channels. For example, in the gluon-gluon scattering channel there are three partonic 
subprocesses contributing to the double real correction: $gg\to gggg$, $gg\to q\bar{q}gg$ and 
 $gg\to q\bar{q}q\bar{q}$; we include the contributions which are leading in the number of 
 colours, $N_{c}$, to all these subprocesses. In practice this amounts to calculating the
 $N_{c}^{2}$, $N_{c}\nf$ and $\nf^2$ corrections to all LO subprocesses, where $\nf$ 
 is the number of light quark flavours. We include the full LO and NLO coefficients in this
 calculation but note that retaining only the leading colour correction to all partonic
 subprocesses at NLO gives the full result to within a few percent across all distributions.
 The analogous subleading colour contributions at NNLO are expected to be small
 and we do not include them in this study. To support this assumption we note that the
  subleading colour NNLO contribution for pure gluon
 scattering was presented in a previous study~\cite{ggSLC} and found to be negligible.
 We construct subtraction terms to regulate all IR
 divergences in the phase space integrals and cancel all explicit poles in the dimensional
 regularization parameter, $\epsilon=(4-d)/2$, the details of which for the antenna subtraction
 method can be found in~\cite{antenna1,gluonsonly,ggSLC}. The IR finite cross section at NNLO is then integrated numerically
 in four dimensions over the appropriate two-, three- or four-parton massless phase space to
 yield the final result.

\begin{figure}[t]
  \centering
    \includegraphics[width=0.5\textwidth]{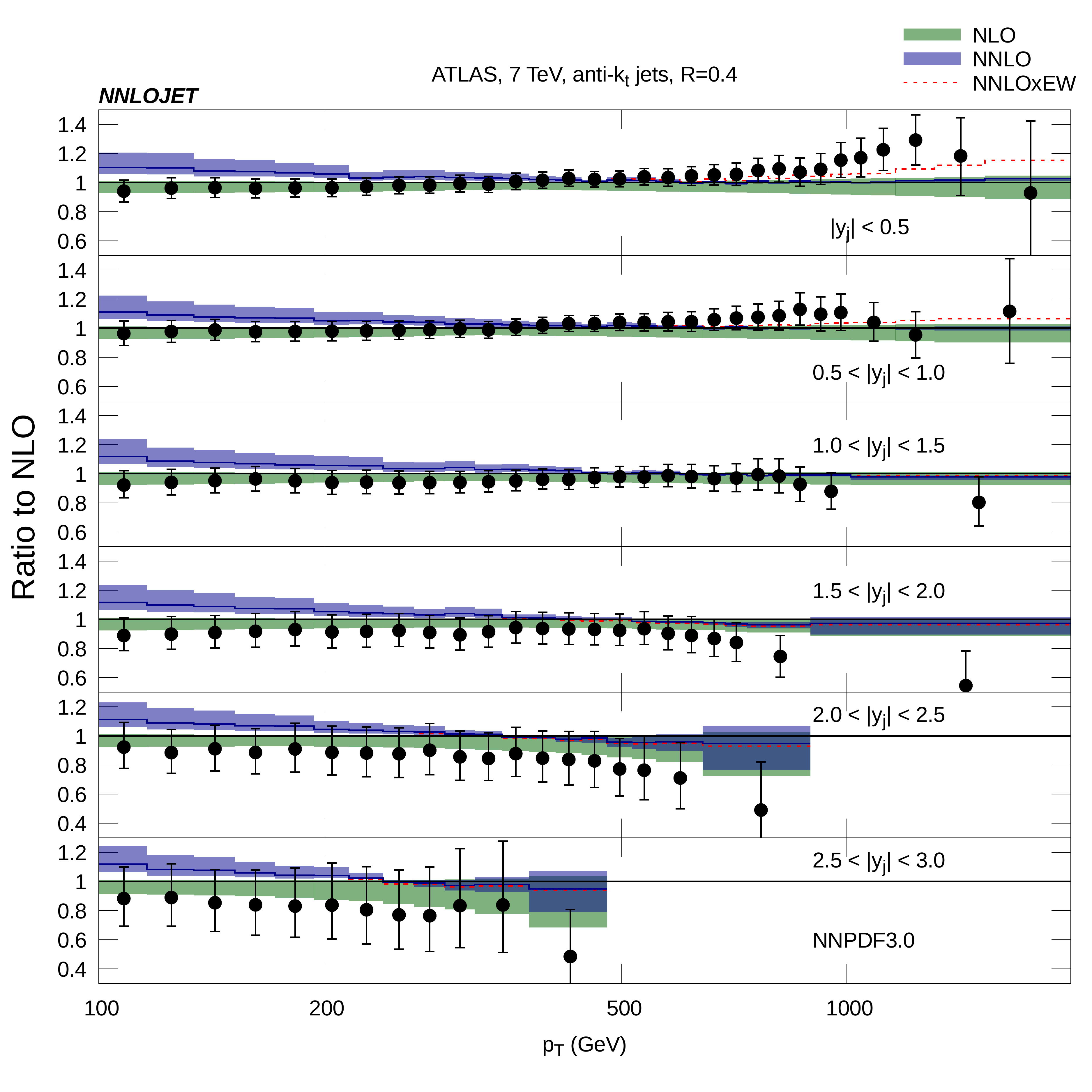}
  \caption{Double-differential inclusive jet cross-sections measurement by ATLAS~\cite{atlasjet2} 
  and NNLO perturbative QCD predictions as a function of the jet $p_T$ in slices of
rapidity, for anti-$k_T$ jets with R = 0.4 normalized to the NLO result. The shaded bands represent the scale
uncertainty of the theory predictions obtained by varying $\mu_R$ and $\mu_F$ as described in the text. The
red dashed line displays the NNLO/NLO ratio corrected multiplicatively for electroweak corrections~\cite{eweak}.}
  \label{fig:dataratio}
\end{figure}
 
In Fig.~\ref{fig:dataratio} we present the results for the double-differential inclusive jet cross section at NLO and NNLO,
normalized to the NLO theoretical prediction to emphasize the impact of the NNLO correction to the NLO result.
The collider setup is proton-proton collisions at a centre of mass energy of 
$\sqrt{s}=$ 7 TeV where the jets are reconstructed using the anti-$k_{T}$ jet algorithm~\cite{antiKT} with $R=0.4$.
We use the NNPDF3.0 NNLO PDF set~\cite{NNPDF} with $\alpha_s(M_Z^2)=0.118$ throughout this paper for LO, NLO and NNLO
predictions to emphasise the behaviour of the higher order coefficient functions at each perturbative order.
 By default we set the renormalization and factorization scales $\mu_R=\mu_F=p_{T1}$, where 
 $p_{T1}$ is the $p_T$ of the leading jet in each event. To obtain the
 scale uncertainty of the theory prediction we vary both scales independently by a factor of 1/2 and 2 with the constraint 
$1/2 \le  \mu_R/\mu_F \le 2$. We find that the NNLO coefficient has a moderate positive effect on the cross section, $10\%$ 
at low $p_{T}$ across all rapidity slices relative to NLO. This is significant because it is precisely in this 
region where the majority of the cross section lies, especially in the central rapidity slices, and it is where we observe 
the largest NNLO effects. At higher $p_{T}$ we see that the relative size of the NNLO correction to NLO decreases 
to the 1-2\% level and so the perturbative series converges rapidly.

\begin{figure}[t]
  \centering
    \includegraphics[width=0.5\textwidth]{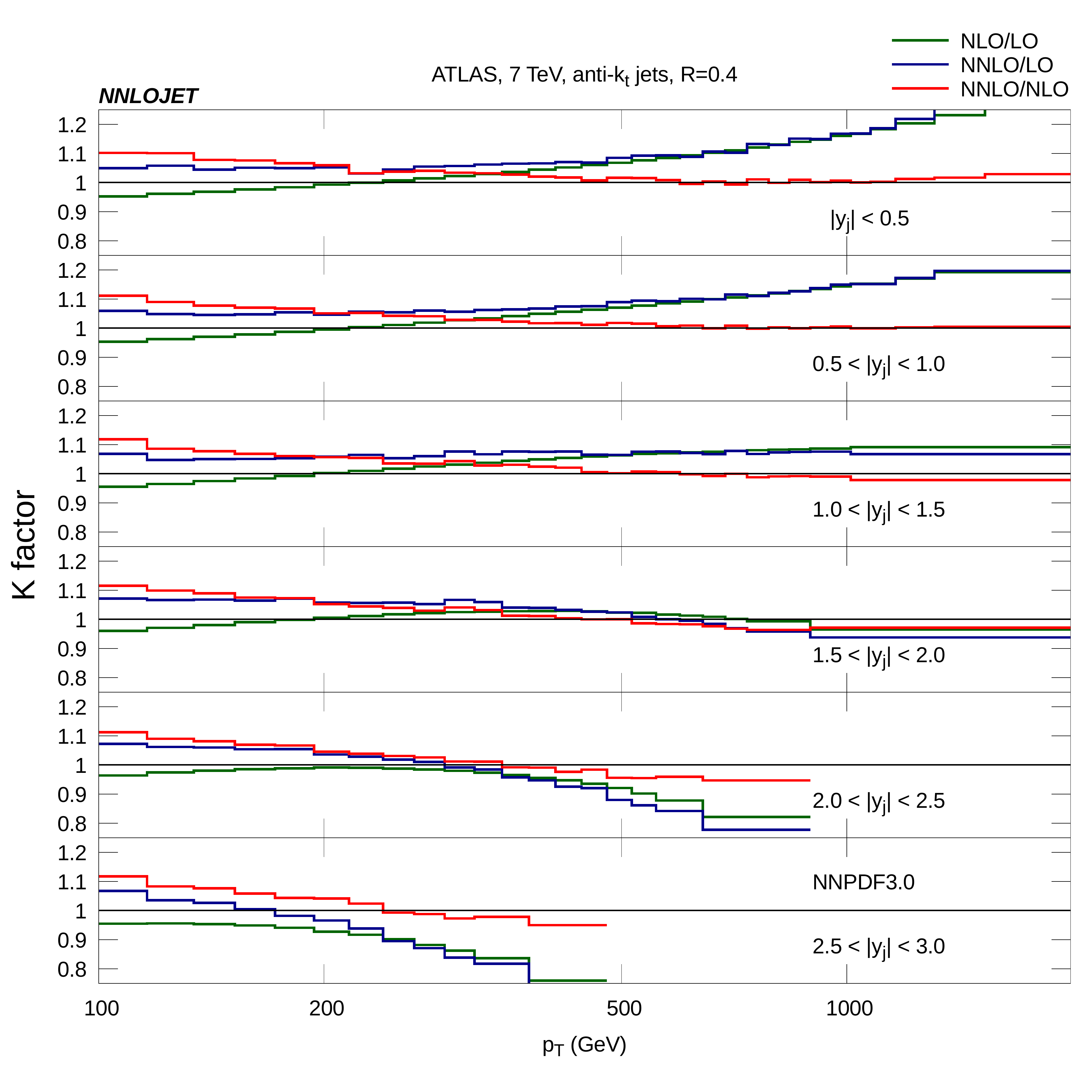}
  \caption{NLO and NNLO $k$-factors for jet production at $\sqrt{s}=7$ TeV. The lines correspond to the double differential
   $k$-factors (ratios of perturbative predictions
  in the perturbative expansion) for $p_T > 100$~GeV and across six rapidity $|y|$ slices.}
  \label{fig:kfact}
\end{figure}

 \begin{figure*}[t]
  \centering
    \includegraphics[width=5.91cm]{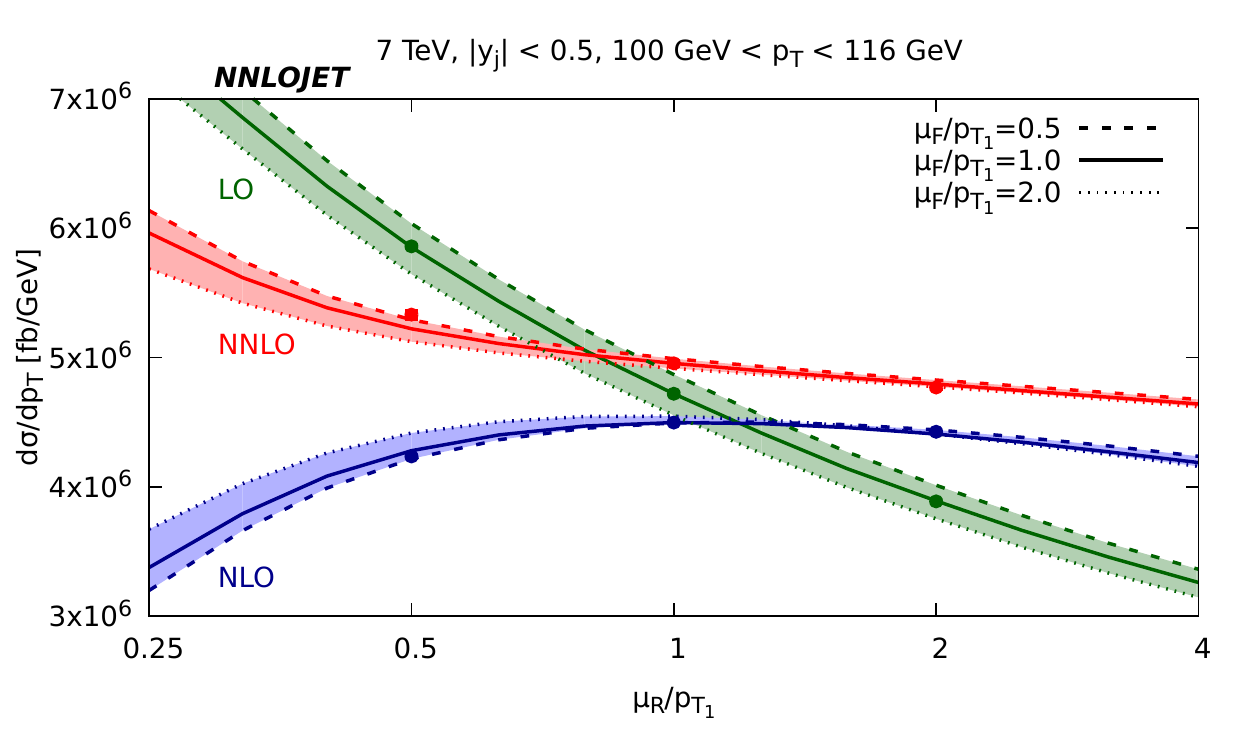}
    \includegraphics[width=5.91cm]{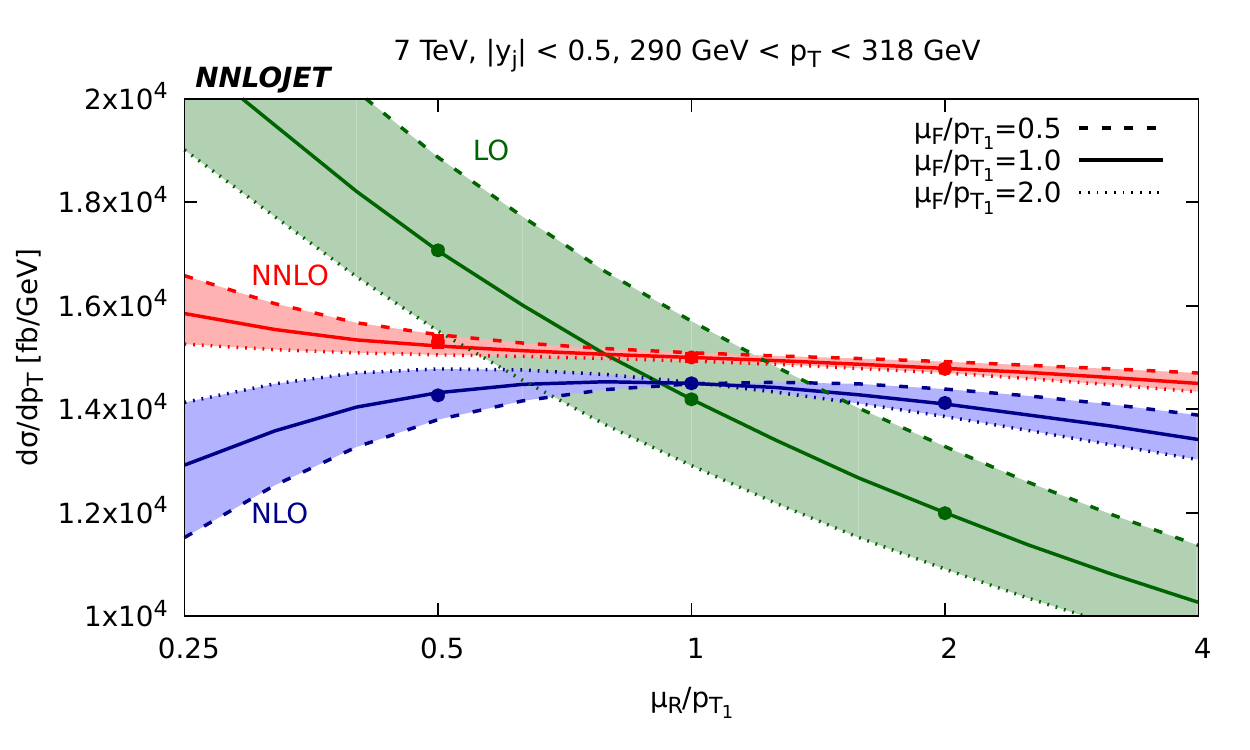}
    \includegraphics[width=5.91cm]{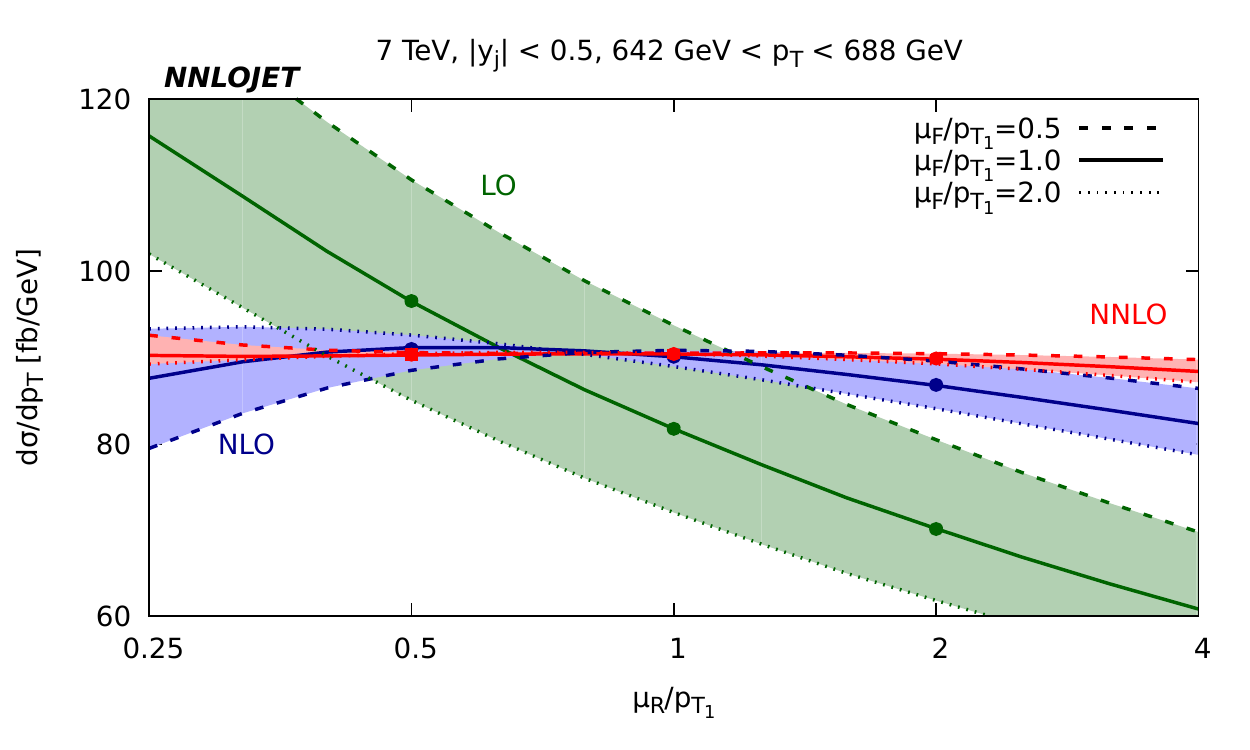}
  \caption{The single inclusive cross section in three $p_{T}$ bins: (left) 100-116 GeV, (centre) 290-318 GeV, (right) 642-688 GeV, plotted against 
  the normalized scale choice $\mu_{R}/p_{T_{1}}$. Points represent the LO, NLO and NNLO cross section as computed by
  \NNLOJET at $\mu_{R}/p_{T_{1}}=0.5,1,2$ and $\mu_{F}/p_{T_{1}}=1$. The solid lines represent the Renormalization Group Equation (RGE) 
  solution for the scale variation, computed
  with $p_{T}$ as the evolution parameter and $\mu_{F}/p_{T_{1}}=1$. Long and short dashed lines represent the same quantities evaluated with 
  $\mu_{F}/p_{T_{1}}=0.5$ and $\mu_{F}/p_{T_{1}}=2$ respectively.}
  \label{fig:scale}
\end{figure*}

Given that we see a moderate NNLO correction to the NLO prediction in the region where the bulk of the cross section lies,
it is instructive to compare to the available data. The data points in Fig.~\ref{fig:dataratio} represent the ATLAS data for an 
integrated luminosity of 4.5 fb$^{-1}$~\cite{atlasjet2}, normalized to the NLO prediction. We do not include non-perturbative
effects in our predictions; they are quantified in~\cite{atlasjet2} and found to be a 2\% effect in the lowest $p_{T}$ bin
and at most a 1\% effect in all other bins (although the quoted uncertainty on the non-perturbative corrections can
be as high as 9\% for the lowest $p_{T}$ bin). The electroweak corrections computed in~\cite{eweak}, are applied multiplicatively to
the QCD calculation for central scale choice using the information provided in~\cite{atlasjet2} and the total is displayed as the red dashed line. 
The electroweak effects are small to moderate for $p_{T}>1$~TeV for central rapidities but otherwise negligible.
We observe that the data is described
very accurately by the NLO prediction, particularly at low to moderate $p_{T}$, whilst the NNLO prediction shows some 
tension with the data in the same region.  

The potential for the NNLO correction to change the shape of the distribution relative to NLO can be seen 
clearly in Fig.~\ref{fig:kfact} where we show the $k$-factors for NLO/LO, NNLO/NLO and NNLO/LO as 
a function of $p_{T}$ in six rapidity slices. In the central rapidity slices we observe that at low $p_{T}$ 
the NLO correction acts negatively relative to LO and then grows to a moderate positive correction at 
high $p_{T}$. In contrast, the NNLO correction acts positively at low $p_{T}$ and decreases to a small effect
at high $p_{T}$. The aggregate effect is shown in the NNLO/LO curve which is the result of a partial
 cancellation between NLO and NNLO at low $p_{T}$, reversing the negative NLO contribution to give a 
 positive total correction, and largely follows the NLO curve at high $p_{T}$. In the 
  region, $|y|> 1.5$, we observe that the NLO correction is once again negative at low $p_{T}$
but does not grow as strongly as in central regions, and is indeed negative at high $p_{T}$ for the most forward
slices. Relative to NLO, the NNLO corrections are again positive and moderate at low $p_{T}$, decreasing in size at high $p_{T}$,
such that the total effect is a positive correction at low $p_{T}$ decreasing to a negative effect at high $p_{T}$ in the most forward region.
\footnote{We observe that the behaviour of the NNLO correction at high $p_{T}$ appears to contradict that predicted by the available
approximate methods based on threshold resummation \cite{threshold1,threshold2}. It should be noted that any direct
comparison between the exact and approximate predictions is limited due to the difference in PDF set, scale choice and experimental
setup used for the published threshold predictions.}

Aside from the size and shape of the NNLO corrections, an interesting feature of Fig.~\ref{fig:dataratio} is the scale
dependence at NLO and NNLO, represented by the thickness of the bands. At high $p_{T}$, especially 
in the central region, the NNLO scale dependence is dramatically reduced and the NNLO band lies firmly within the NLO band.
The situation is once again different at low $p_{T}$ where we observe an appreciable scale variation on the NNLO
calculation, in some places even larger than the NLO scale variation, and the bands do not fully overlap.

This behaviour is unexpected and so in Fig.~\ref{fig:scale} we analyse the scale variation in more detail. We select three
$p_{T}$ bins at low (100-116 GeV), intermediate (290-318 GeV) and high (642-688 GeV) $p_{T}$ in 
the central region $|y|<0.5$ and display the cross section as a function of $\mu_{R}/p_{T_{1}}$.
The points represent the cross section as calculated at LO, NLO and NNLO,
evaluated at $\mu_{R}/p_{T_{1}}\in\{0.5,1,2\}$ with $\mu_{F}/p_{T_{1}}=1$. The solid line
is the Renormalization Group Equation (RGE) prediction for the scale variation
starting from the cross section computed with $\mu_{R}/p_{T_{1}}=\mu_{F}/p_{T_{1}}=1$.
To be fully consistent, the RGE evolution variable should be $p_{T_{1}}$, however starting with the distribution
${\rm{d}}\sigma/{\rm{d}}p_{T}$, all information on $p_{T_{1}}$ is lost and therefore we use 
$p_{T}$ to approximate $p_{T_{1}}$ as the RGE evolution variable~\footnote{For Born
kinematics (appropriate to LO, virtual NLO and double virtual NNLO contributions) and the hardest jet in any real NLO or double real NNLO event 
$p_{T}=p_{T_{1}}$ exactly. For the subleading jets in real emission contributions this only approximately holds,
 although the cross section for events with a large hierarchy in $p_{T}$ between leading and subleading jets is a small contribution to the total. 
We therefore expect, and observe, the difference between using $p_{T_{1}}$ and $p_{T}$ as the RGE evolution variable to be small.}.
Nevertheless, the difference in the evolution is small even at low $p_{T}$ 
and we include the RGE lines to aid the discussion of the scale variation in each bin. The long and short dashed lines are obtained using 
 $\mu_{F}/p_{T_{1}}=0.5$ and $\mu_{F}/p_{T_{1}}=2$ respectively.
 
 From the left pane in Fig.~\ref{fig:scale} we observe that at LO the scale variation is a monotonically decreasing function
 with $p_{T}$. At NLO the picture is quite different; the shape of the RGE curve turns over at approximately the central scale choice.
 The consequence of this behaviour is that the scale band is asymmetric, with the central scale being located at the upper edge of the band,
 as can be seen in Fig.~\ref{fig:dataratio} where the NLO scale band lies almost entirely below one. The overall variation is also
 relatively small at low $p_{T}$, ($<5\%$), which is linked to the smallness of the NLO coefficient,
  ($\sim4\%$ of LO), as displayed in Fig.~\ref{fig:kfact}. The NNLO curve is monotonically decreasing, leading 
  to a more symmetric band and the overall variation is significant, largely reflecting the size of the NNLO correction. In the central
  and right panes of Fig.~\ref{fig:scale} we observe a similar behaviour; the NLO curve is once again turning over, yielding an asymmetric scale band in
  Fig.~\ref{fig:dataratio}. The NNLO curves are monotonically decreasing and show diminishing scale variation with increasing
  $p_{T}$ as the NNLO coefficient becomes negligible relative to the moderately large NLO correction, as displayed in Fig.~\ref{fig:kfact}.
  
  To understand these results it is instructive to consider the low and high $p_{T}$ regions separately. At high $p_{T}$, the
  NLO correction is moderate and positive, the NNLO correction is small and reduces the scale variation significantly
  relative to NLO with the NNLO scale band lying within the NLO band. At low $p_{T}$ the behaviour is quite 
  different; we observe that the NLO correction is small and generally negative whereas the NNLO correction is moderate and 
  positive. The relative size of the NLO and NNLO corrections leads to a significant NNLO scale band which largely does not 
  overlap with the NLO band.
  
  A possible explanation may be found in the comparison to data in Fig.~\ref{fig:dataratio}. The data 
 appears to be consistent with the NLO prediction
  when using the NNLO NNPDF3.0 PDF set. 
  This PDF set is fitted to Tevatron and LHC jet data at NNLO despite the exact calculation for jet production being reported for the
  first time in this letter. 
  In the approximate NNLO PDF fit,  a restricted data set at high $p_T$ was used in conjunction with an approximation to the NNLO coefficient functions.  This procedure has a small influence on the low $p_T$ region which was included in the NLO PDF fit and in fact we find that the NLO prediction using the NLO PDF set gives an almost identical result to that shown in Fig.~\ref{fig:dataratio} obtained with NNLO PDFs. However, at low $p_T$ we see that the exact NNLO correction is not small and the NNLO PDF fit potentially underestimates the effect of the NNLO contribution in that region. A detailed study of the effects of the single jet inclusive datasets and NNLO theory predictions on PDF fits is required for more substantive conclusions.
       
  We have presented the first calculation of the single jet inclusive cross section including all partonic subprocesses at NNLO for the LHC 
  using the antenna subtraction method, implemented in the parton-level event generator \NNLOJET and compared to ATLAS data. 
  We find that the NNLO corrections
  are moderate at low $p_{T}$ and change the shape of the distributions relative to NLO. At high $p_{T}$ the corrections are smaller
  and we see a dramatic reduction in the scale variation. The fact that the NNLO corrections move the theoretical prediction away from
  the available data suggests that this calculation may have an appreciable impact when included in refitting the PDF used in this
  study to jet data. We anticipate that this calculation will open the door for precision phenomenology using LHC jet data, including
  studies of scale choice, jet shape, cone size and different PDF sets. 
  
  The authors thank Xuan Chen, Juan Cruz-Martinez, Aude Gehrmann-De Ridder, Thomas Gehrmann, Alexander Huss, 
  Tom Morgan and Jan Niehues for useful discussions and their many contributions to the \NNLOJET code.
   We thank Johannes Bluemlein, Joey Huston and Klaus Rabbertz for their useful comments on the draft manuscript.
   We gratefully acknowledge the assistance provided by Jeppe Andersen utilizing the computing resources 
   provided by the WLCG through the GridPP Collaboration. 
   This research was supported in part by the UK Science and Technology Facilities Council, in part by the Research 
   Executive Agency (REA) of the European Union under the Grant Agreement PITN-GA-2012-316704 (``HiggsTools'')
    and the ERC Advanced Grant MC@NNLO (340983).


\begin{thebibliography}{99}

\bibitem{disbook}
  R.~Devenish and A.~Cooper-Sarkar,
 {\it Deep inelastic scattering}, Oxford University Press (Oxford, 2004).
    
 \bibitem{disjet}
  P.~Newman and M.~Wing,
  Rev.\ Mod.\ Phys.\  {\bf 86} (2014)  1037.


\bibitem{cdf} 
  A.~Abulencia {\it et al.} [CDF Collaboration],
  Phys.\ Rev.\ D {\bf 75}, 092006 (2007)
  Erratum: [Phys.\ Rev.\ D {\bf 75}, 119901 (2007)].
  
\bibitem{d0} 
  V.~M.~Abazov {\it et al.} [D0 Collaboration],
  Phys.\ Rev.\ D {\bf 85}, 052006 (2012)
  
  \bibitem{atlasjet}
  G.~Aad {\it et al.} [ATLAS Collaboration],
  Eur.\ Phys.\ J.\ C {\bf 71}, 1512 (2011);
  Eur.\ Phys.\ J.\ C {\bf 73}, 2509 (2013).
  
\bibitem{atlasjet2} 
  G.~Aad {\it et al.} [ATLAS Collaboration],
  JHEP {\bf 1502}, 153 (2015)
  Erratum: [JHEP {\bf 1509}, 141 (2015)].

\bibitem{cmsjet}
  V.~Khachatryan {\it et al.} [CMS Collaboration],
  Eur.\ Phys.\ J.\ C {\bf 76}, 451 (2016);
  JHEP {\bf 1206}, 036 (2012);
  Phys.\ Rev.\ Lett.\  {\bf 107}, 132001 (2011).
    
\bibitem{eks}
S.~D. Ellis, Z.~Kunszt and D.~E. Soper, {Phys.\ Rev.\ Lett.} {\bf 69}
  (1992) 1496.

\bibitem{jetrad}
W.~T. Giele, E.~W.~N. Glover and D.~A. Kosower,  {Phys.\
  Rev.\ Lett.} {\bf 73} (1994) 2019.

\bibitem{nlojet1}
Z.~Nagy, {Phys.\ Rev.\ Lett.} {\bf 88} (2002) 122003;
{Phys.\ Rev.} {\bf D68} (2003) 094002.

\bibitem{powheg2j}
S.~Alioli, K.~Hamilton, P.~Nason, C.~Oleari and E.~Re,  
{JHEP} {\bf 1104} (2011) 081.

\bibitem{meks}
  J.~Gao, Z.~Liang, D.~E.~Soper, H.~L.~Lai, P.~M.~Nadolsky and C.-P.~Yuan,
  Comput.\ Phys.\ Commun.\  {\bf 184}, 1626 (2013).
  
  \bibitem{partial}
   A.~Gehrmann-De Ridder, T.~Gehrmann, E.~W.~N.~Glover and J.~Pires,
  Phys.\ Rev.\ Lett.\  {\bf 110}, no. 16, 162003 (2013);
  J.~Currie, A.~Gehrmann-De Ridder, T.~Gehrmann, N.~Glover, J.~Pires and S.~Wells,
  arXiv:1407.5558 [hep-ph].

\bibitem{ABM12} 
  S.~Alekhin, J.~Blumlein and S.~Moch,
  Phys.\ Rev.\ D {\bf 89}, 054028 (2014)

\bibitem{NNPDF}
  R.~D.~Ball {\it et al.} [NNPDF Collaboration],
  JHEP {\bf 1504}, 040 (2015).
\bibitem{MMHT} 
  L.~A.~Harland-Lang, A.~D.~Martin, P.~Motylinski and R.~S.~Thorne,
  Eur.\ Phys.\ J.\ C {\bf 75}, 204 (2015).
  \bibitem{CT14}
  S.~Dulat {\it et al.},
  Phys.\ Rev.\ D {\bf 93}, 033006 (2016).
  
  
  \bibitem{real}
M.~L. Mangano and S.~J. Parke,  {Phys.\ Rept.} {\bf 200} (1991) 301.


\bibitem{fiveg1l} 
  Z.~Bern, L.~J.~Dixon and D.~A.~Kosower,
  Phys.\ Rev.\ Lett.\  {\bf 70}, 2677 (1993).


\bibitem{2q3g1l} 
  Z.~Bern, L.~J.~Dixon and D.~A.~Kosower,
  Nucl.\ Phys.\ B {\bf 437}, 259 (1995).


\bibitem{4q1g1l} 
  Z.~Kunszt, A.~Signer and Z.~Trocsanyi,
  Phys.\ Lett.\ B {\bf 336}, 529 (1994).

\bibitem{twol}
E.~W.~N. Glover, C.~Oleari and M.~E. Tejeda-Yeomans, {Nucl.\ Phys.} 
B {\bf 605} (2001) 467; 
E.~W.~N. Glover and M.~Tejeda-Yeomans, {JHEP} {\bf 0105} (2001) 010;
Z.~Bern, A.~De Freitas and L.~J.~Dixon,
  {JHEP} {\bf 0203} (2002) 018.
  
\bibitem{quarkgluon2l} 
  C.~Anastasiou, E.~W.~N.~Glover, C.~Oleari and M.~E.~Tejeda-Yeomans,
  Nucl.\ Phys.\ B {\bf 605}, 486 (2001);
  Phys.\ Lett.\ B {\bf 506}, 59 (2001);
  JHEP {\bf 0306}, 033 (2003);
  Z.~Bern, A.~De Freitas and L.~J.~Dixon,
  JHEP {\bf 0306}, 028 (2003).
  Erratum: [JHEP {\bf 1404}, 112 (2014)].

  
  
  
\bibitem{fourq2l} 
  C.~Anastasiou, E.~W.~N.~Glover, C.~Oleari and M.~E.~Tejeda-Yeomans,
  Nucl.\ Phys.\ B {\bf 601}, 318 (2001);
  Nucl.\ Phys.\ B {\bf 601}, 341 (2001);
  Phys.\ Lett.\ B {\bf 506}, 59 (2001);
  A.~De Freitas and Z.~Bern,
  JHEP {\bf 0409}, 039 (2004).
  
  \bibitem{antenna1}
 A.~Gehrmann-De Ridder, T.~Gehrmann and E.~W.~N.~Glover,
  JHEP {\bf 0509}, 056 (2005);
 J.~Currie, E.~W.~N.~Glover and S.~Wells,
  JHEP {\bf 1304}, 066 (2013).
    
\bibitem{antenna2}
  A.~Daleo, T.~Gehrmann and D.~Maitre,
  JHEP {\bf 0704}, 016 (2007);
 A.~Daleo, A.~Gehrmann-De Ridder, T.~Gehrmann and G.~Luisoni,
  JHEP {\bf 1001}, 118 (2010);
  R.~Boughezal, A.~Gehrmann-De Ridder and M.~Ritzmann,
  JHEP {\bf 1102}, 098 (2011);
   T.~Gehrmann and P.~F.~Monni,
  JHEP {\bf 1112}, 049 (2011);
  A.~Gehrmann-De Ridder, T.~Gehrmann and M.~Ritzmann,
  JHEP {\bf 1210}, 047 (2012).


 \bibitem{qtsub}
  S.~Catani and M.~Grazzini,
  Phys.\ Rev.\ Lett.\  {\bf 98} (2007) 222002.
  
  \bibitem{njettiness}
  R.~Boughezal, C.~Focke, X.~Liu and F.~Petriello,
  Phys.\ Rev.\ Lett.\  {\bf 115} (2015)  062002;
    R.~Boughezal, X.~Liu and F.~Petriello,
  Phys.\ Rev.\ D {\bf 91} (2015)  094035;
    J.~Gaunt, M.~Stahlhofen, F.~J.~Tackmann and J.~R.~Walsh,
  JHEP {\bf 1509} (2015) 058.
  
   \bibitem{stripper}
  M.~Czakon,
  Phys.\ Lett.\ B {\bf 693} (2010) 259;
   R.~Boughezal, K.~Melnikov and F.~Petriello,
  Phys.\ Rev.\ D {\bf 85} (2012) 034025.


 \bibitem{secdec}
  T.~Binoth and G.~Heinrich,
  Nucl.\ Phys.\ B {\bf 693} (2004) 134;
  C.~Anastasiou, K.~Melnikov and F.~Petriello,
  Phys.\ Rev.\ D {\bf 69} (2004) 076010.

    
\bibitem{projecttoborn} 
  M.~Cacciari, F.~A.~Dreyer, A.~Karlberg, G.~P.~Salam and G.~Zanderighi,
  Phys.\ Rev.\ Lett.\  {\bf 115}, 082002 (2015)


\bibitem{zjet} 
  A.~Gehrmann-De Ridder, T.~Gehrmann, E.~W.~N.~Glover, A.~Huss and T.~A.~Morgan,
  Phys.\ Rev.\ Lett.\  {\bf 117}, 022001 (2016).

\bibitem{zjetproc} 
  A.~Gehrmann-De Ridder, T.~Gehrmann, E.~W.~N.~Glover, A.~Huss and T.~A.~Morgan,
  arXiv:1601.04569 [hep-ph].

\bibitem{ggSLC}
 J.~Currie, A.~Gehrmann-De Ridder, E.~W.~N.~Glover and J.~Pires,
  JHEP {\bf 1401}, 110 (2014).

  \bibitem{antiKT}
    M.~Cacciari, G.~P.~Salam and G.~Soyez,
  JHEP {\bf 0804}, 063 (2008).
  
  \bibitem{gluonsonly}
 E.~W.~N.~Glover and J.~Pires,
  JHEP {\bf 1006}, 096 (2010);
  A.~Gehrmann-De Ridder, E.~W.~N.~Glover and J.~Pires,
  JHEP {\bf 1202}, 141 (2012);
  A.~Gehrmann-De Ridder, T.~Gehrmann, E.~W.~N.~Glover and J.~Pires,
  JHEP {\bf 1302}, 026 (2013).

\bibitem{eweak} 
  S.~Dittmaier, A.~Huss and C.~Speckner,
  JHEP {\bf 1211}, 095 (2012).
  

\bibitem{threshold1} 
M.~C.~Kumar and S.~O.~Moch, 
Phys.\ Lett.\ B {\bf 730}, 122 (2014).

\bibitem{threshold2} 
D.~de Florian, P.~Hinderer, A.~Mukherjee,
F.~Ringer and W.~Vogelsang, 
Phys.\ Rev.\ Lett.\ {\bf 112}, 082001 (2014).
    
\end{thebibliography}
\end{document}